# Optimal Service Provisioning in IoT Fog-based Environment for QoS-aware Delay-sensitive Application

Soroush Hashemifar, Amir Rajabzadeh[*]

*Department of Electrical and Computer Engineering, Razi University, Kermanshah, Iran*

**Abstract**

This paper addresses the escalating challenges posed by the ever-increasing data volume, velocity, and the demand for low-latency applications, driven by the proliferation of smart devices and Internet of Things (IoT) applications. To mitigate service delay and enhance Quality of Service (QoS), we introduce a hybrid optimization of Particle Swarm (PSO) and Chemical Reaction (CRO) to improve service delay in FogPlan, an offline framework that prioritizes QoS and enables dynamic fog service deployment. The method optimizes fog service allocation based on incoming traffic to each fog node, formulating it as an Integer Non-Linear Programming (INLP) problem, considering various service attributes and costs. Our proposed algorithm aims to minimize service delay and QoS degradation. The evaluation using real MAWI Working Group traffic data demonstrates a substantial 29.34% reduction in service delay, a 66.02% decrease in service costs, and a noteworthy 50.15% reduction in delay violations compared to the FogPlan framework.

## 1 Introduction

In recent years, the Internet of Things (IoT) has emerged to facilitate various applications, such as smart homes, wearable devices, smart vehicles, and connected healthcare. Consequently, the amount of data generated by IoT devices is increasing massively. By looking to the future, a new forecast from International Data Corporation (IDC) estimates that there will be 41.6 billion connected IoT devices generating 79.4 zettabytes (ZB) of data in 2025 [1]. Organizations and businesses are intensely investing in IoT solutions to capture and store massive amounts of data to help them get insights. These insights would help them develop new products, accelerate decision-making processes, improve processes, reduce costs, and detect anomalies before they happen. The challenge lies in their ability to use data streams and react immediately to critical.

---

[*] Corresponding author. Tel. +98 833 434 34 10; URL: https://razi.ac.ir/en/~rajabzadeh (A. Rajabzadeh), Email: rajabzadeh@razi.ac.ir

Knowing that edge devices usually do not have the computational and storage capabilities for local processing of generated data, the traditional approach is to transfer data to the cloud, which has massive storage and processing resources. This approach seems to be costly concerning latency, network bandwidth, storage, and energy use. The cloud does not satisfy delay and Quality of Service (QoS) requirements of delay-sensitive IoT applications due to its long distance from different terminal devices and even privacy issues [2].

Taking computing, storage, and networking resources closer to data sources, *i.e.*, IoT devices, not only addresses the problems mentioned earlier [3] but also enhances user experience [4]. Fog computing [5], as a geo-distributed computing paradigm, can put this idea into practice resulting in a tangible reduction in delay, bandwidth usage, and energy consumption, and improving service quality. Notice that it is not replacing cloud systems; it is just a complementary computing paradigm. These fog services are provided as virtual machines and containers. While a fog computing environment presents various challenges, we will specifically concentrate on two of these difficulties. Service provisioning is a fundamental problem in fog computing due to fog nodes' limited computing capacity and the rapid development of delay-sensitive applications. On the other hand, fog can increase service delay and energy consumption if the service provisioning on cloud servers or fog nodes is not well-balanced, wildly when each service's resource usage varies over time. A dynamic mechanism for provisioning services is essential for efficiently deploying services to computing resources. According to [6], there exist two families of application services: A service with a strict delay requirement must be deployed on the edge of the fog network, *i.e.*, *delay-sensitive services*, and a delay-tolerant service that can be deployed on other fog nodes or the cloud, *i.e.*, *availability-sensitive services*.

If the fog service is deployed too far from IoT devices, QoS degradation will occur, leading to an increase in transmission delay [7]. The provisioning task can be performed statically. In this case, it is crucial to select an adequate number of services to implement. This will lead to a decrease in resource expenses and help avoid exceeding the specified threshold.

On the other hand, incoming traffic to fog nodes changes over time, which causes changes in services' resource usage [6], so the provisioning strategy can not adapt itself to these changes. A fog service placement that is dynamic and allows for the deployment and release of fog services in a dynamic manner can effectively solve the problem that has been mentioned.

This paper formulated the fog service placement process as an optimization problem using Integer Non-Linear Programming (INLP). Since dynamic fog service placement is an NP-hard problem, a meta-heuristic algorithm is proposed based on hybrid binary particle swarm optimization (BPSO) [8] and chemical reaction optimization (CRO) [9]. The algorithm aims to minimize the average service delay and reduce the overall QoS loss concerning resources' computing and storage capacities. We have formulated a cost function that encompasses the integration of storage, processing, deployment, and communication costs, as well as penalties for delay violation and wrong provisions in order to attain this objective. Since meta-heuristic algorithms are slow, we run the algorithm offline, which allows us to determine the optimal placement for a while. During this period, the Fog Service Controller (FSC) provisions services according to the last placement until the algorithm determines a new placement. To make this approach online, one can predict the requests and provision them before running the algorithm in FSC. We have left this as future work and described it in detail in the last section of the paper.

To evaluate the HPCDF (Hybrid PSO-CRO Delay-improved for FogPlan), three different experiments are performed to investigate the impact of hyper-parameters of the framework and the hybrid binary PSO-CRO[1] (HBPCRO) algorithm. An extra experiment is conducted to compare the optimality of the proposed framework with the methods of the FogPlan framework [6]. All the experiments are performed on real-world traffic traces from the MAWI[2] Working Group, which contains 2 hours of traffic split into chunks of 15-minute intervals.

The main contributions of this paper are summarized as follows:

- *QoS-aware framework*: We have included QoS violation as a part of the cost function, and a penalty is assigned to it. The meta-heuristic algorithm aims to keep QoS violation below the customer-expected threshold for a certain fraction of the time.
- *Optimal service provisioning*: We have proposed a hybrid binary PSO-CRO algorithm for fog service provisioning. This hybrid meta-heuristic algorithm has been proven to find the optimal optimization solution.
- *Minimum service information is required*: Thus, we consider only the aggregated incoming service traffic to fog nodes; no additional service information is required for the provisioning task.
- *Offline approach*: Unlike the state-of-the-art approaches, HPCDF performs offline. Hence, it has enough time to explore the search space. It also adapts itself to changes that occur on incoming traffic to fog nodes.

The rest of this paper is organized as follows. Section 2 presents a brief of state-of-the-art methods that have been proposed for fog service provisioning. The service provisioning problem is formulated in Section 3. Then, the proposed approach is described in Section 4. Section 5 provides the evaluation settings and results with numerical simulations. Finally, Sections 6 and 7 give the discussion and conclusion and state the future work.

---

[1] Particle Swarm- Chemical Reaction Optimization
[2] Measurement and Analysis on the WIDE Internet

## 2   Related work

As fog computing has emerged as a relatively new paradigm, several open problems exist within it. One of these problems is optimizing the placement of tasks on available resources, *i.e.*, service provisioning problems. Many studies have been done on solving task scheduling in heterogeneous environments. Yousefpour et al. have put forward an application-independent framework for provisioning stateless service in a hybrid cloud-fog environment called FogPlan [6]. It has also proposed two greedy online approaches. MinViol deploys services with high traffic and releases them if their traffic decreases. On the other hand, MinCost deploys services to reduce costs and increase revenue. Each of these algorithms is invoked periodically to update the placement of services. FogPlan does not need any information about the mobility of users or specs of IoT devices. The experiments show that MinCost performs faster than MinViol, but MinViol has lower service delay and delay violation. Lera et al. proposed a Graph-Partition-based strategy for deployment of applications and services [10]. Their proposed method maximizes service availability and QoS levels by minimizing network latency between unrelated services. First, they map the application to the fog community using a greedy first-fit decreasing approach to increase service availability. Deployed application services are then assigned to fog nodes within that community to prevent deadline violations. If there are no resources accessible on any service, the application will be deployed to the cloud. Donassolo et al. proposed a service provisioning mechanism based on a divide-and-conquer approach in foggy environments [11]. The goal is to minimize deployment costs. First, decompose the problem into multiple solution components. Each solution component then determines its placement using a greedy strategy concerning requirements.

Huang et al. present a formulation of the general problem of multi-replica service placement and data flow management [12]. They model this problem as a multi-objective scheduling problem. This method was developed based on Pareto-ACO (P-ACO). The fog service placement problem was introduced by Natesha et al., who applied a genetic algorithm that was based on the elite approach to address it [13]. Although they optimized energy consumption, service time, and cost, the proposed scheme did not consider the fog layer's resource usage. Jafari et al. used a meta-heuristic approach to optimize latency and power consumption in the Cloud-Fog-IoTecosystem simultaneously [14]. The authors present two meta-heuristic approaches, including the Bees Algorithm (BA) and the Non-Dominant Genetic Reordering Algorithm (NSGA-II), to solve the service placement problem (SPP) in the form of a multi-objective problem. BA and NSGA-II are combined with a robust differential evolution method called Minimax Differential Evolution (MMDE) to improve convergence. Simulations demonstrate that NSGA-II outperforms BA in terms of both latency and power. Partial Offloading to Multiple Helper (POMH) is a task offloading technique that provides parallel computation of tasks and improves task latency. Tasks are split into subtasks and transferred to different fog nodes in parallel, thus reducing the total computation time [15]. They performed POMH task offloading using horizontal task offloading to adjacent fog nodes and vertical offloading to the cloud. They proposed a broad framework for minimizing service delivery delays through adaptive task offload mechanisms.

Wu et al. deal with SPP, meeting different QoS requirements regarding cost and deadlines [16]. SPP is formulated as a multi-objective optimization problem with tradeoffs between other objectives while maintaining a set of Pareto optimizations. They developed a genetic algorithm that is a meta-heuristic method, which is built upon a common parallel structure and employs multiple elite operators in order to enhance the solution for the SPP. The proposed scheme is designed with a two-way trust management mechanism to ensure both QoS and reliability simultaneously, in order to address the limitations of fog computing. The state-of-the-art works proposed in this section are summarized in Table 1. By the way, as far as we know, none of the previous research studies have considered service delay, deadline violation, and financial cost in addition to utilization when it comes to providing services more efficiently in the cloud-fog environment.

Table 1. Summary of various related work

| Paper | Year | Objectives | | | | Environment | Method |
|---|---|---|---|---|---|---|---|
| | | Delay | Deadline violation | Service cost | Utilization | | |
| Yousefpour et al. [6] | 2019 | ✓ | ✓ | ✓ | | cloud-fog | Greedy |
| Lera et al. [10] | 2019 | ✓ | ✓ | | | cloud-fog | Greedy |
| Donassolo et al. [11] | 2019 | | | ✓ | | fog | Greedy |
| Huang et al. [12] | 2020 | ✓ | | ✓ | | cloud-fog | Ant Colony |
| Natesha et al. [13] | 2021 | ✓ | | ✓ | | fog | Genetic Algorithm |
| Jafari et al. [14] | 2021 | ✓ | | | | cloud-fog | Bees Algorithm and NSGA-II |
| Tran-Dang et al. [15] | 2021 | ✓ | | | | cloud-fog | Greedy |
| Wu et al. [16] | 2022 | ✓ | ✓ | ✓ | ✓ | fog | Genetic Algorithm |
| Our method | 2022 | ✓ | ✓ | ✓ | ✓ | cloud-fog | HBPCRO |

## 3 System Model and Objective Statement

### 3.1 System Model

In this section, we begin by defining the system model for our proposed algorithm. We establish expressions for queuing delay, service delay, and constraints. We formulate the optimization problem using INLP. The FSC, part of the Fog Service Provider (FSP), allocates resources to user applications based on request volume and available fog and cloud resources. IoT applications consist of interconnected micro-services executed as containers or virtual machines with specific resource requirements. The objective is to place these services on diverse fog nodes and cloud servers, minimizing a cost function while meeting constraints and maximizing QoS by service-level agreements (SLA). Service migration decisions are based on response delays determined by FSC. Figure 1 illustrates interactions in the hybrid fog-cloud environment.

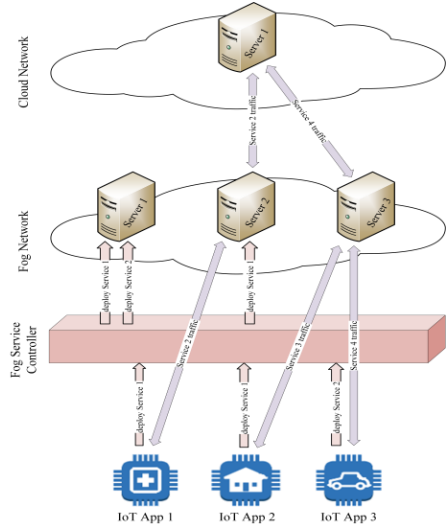
Figure 1. Service deployment in a hybrid fog-cloud environment

In Figure 1, we assumed that each fog node could only process two services simultaneously. In this scenario, the third fog node cannot process the fourth service, which leads to deploying it on the cloud server. On the other hand, the first service's arrived requests are forwarded to newly deployed instances of the first service.

### 3.1.1 Assumptions and Decision Variables

$F$ and $C$ denote fog nodes and cloud servers throughout this paper, respectively, and $S$ represents the services. Moreover, $q_s$ and $th_s$ show the QoS level and service delay threshold, respectively. FSP has to guarantee that delay will be lower than $th_s$ for $q_s$ percentage of the time.

We determine the placement of services on fog nodes and cloud servers at any moment through both fog placement ($P$) and cloud placement ($Q$) matrices as follows:

$$P(s, f, t) = \begin{cases} 1 & \text{if service s is deployed on node f} \\ 0 & \text{otherwise} \end{cases} \quad (1)$$

$$Q(s, k, t) = \begin{cases} 1 & \text{if service s is deployed on server k} \\ 0 & \text{otherwise} \end{cases} \quad (2)$$

Each row of matrices $P$ and $Q$ shows the service number. Each column of matrix $P$ indicates the fog node number, and each column of matrix $Q$ indicates the cloud server number. All notations are shown in Table 2.

*3.1.2 Required processing resources*

The number of instructions that arrive at a fog node per second is defined by the amount of processing capacity that the service has occupied, *i.e.*, $R_{proc}(s)$, as well as on the amount of incoming traffic to the fog node due to service residency, *i.e.*, $T^{fog}(s,f,t)$.

Table 2. Summary of Notations

| Symbol | Definition | Unit | Symbol | Definition | Unit |
|---|---|---|---|---|---|
| $\tau$ | The period between two consecutive runs of the algorithm | s | $M_{proc}^{cloud}(k)$ | Processing capacity of cloud server $k$ | MIPS |
| F | Set of all fog nodes | | $M_{mem}^{cloud}(k)$ | Memory capacity of cloud server $k$ | Byte |
| C | Set of all cloud servers | | $M_{stor}^{cloud}(k)$ | Storage capacity of cloud server $k$ | Byte |
| $th_s$ | Max. delay threshold of service $s$ | ms | $R_{proc}(s)$ | Required processing capacity for service $s$ | MI[3]/request |
| $P(s,f,t)$ | Placement of service $s$ on fog node $f$ at time t | | $R_{mem}(s)$ | Required memory capacity for service $s$ | Byte |
| $Q(s,k,t)$ | Placement of service $s$ on cloud server $k$ at time t | | $R_{stor}(s)$ | Required storage capacity for service $s$ | Byte |
| $d_{service}(s,f,t)$ | Service delay for service $s$ on fog node $f$ at time t | ms | $X$ | Particle position vector | |
| $\psi^{fog}(s,f,t)$ | Arrival instructions of service $s$ to fog node $f$ | MIPS | $V$ | Particle velocity vector | |
| $T^{fog}(s,f,t)$ | Incoming requests rate of service $s$ to fog node $f$ | request/s | $p_{best}$ | The local best particle position vector | |
| $\Gamma^{fog}(s,f,t)$ | Processing capacity that $f$ is dedicated to service $s$ | MIPS | $g_{best}$ | The global best particle position vector | |
| $\psi^{cloud}(s,k,t)$ | Arrival instructions of service $s$ to cloud server $k$ | MIPS | $\omega$ | Inertia coefficient | |
| $\Gamma^{cloud}(s,k,t)$ | Processing capacity that $k$ dedicates to service $s$ | MIPS | $\alpha$ | Cognitive coefficient | |
| $h_s(f)$ | Target cloud server for offloaded service $s$ from $f$ | | $\beta$ | Social coefficient | |
| $\xi(f)$ | Transmission rate between FSC and fog node $f$ | Byte/s | $\gamma$ | Maximum times of local search in PSO | |
| $M_{proc}^{fog}(f)$ | Processing capacity of fog node $f$ | MIPS[4] | $V_{min}$ | Minimum velocity value | |
| $M_{mem}^{fog}(f)$ | Memory capacity of fog node $f$ | Byte | $V_{max}$ | Maximum velocity value | |
| $M_{stor}^{fog}(f)$ | Storage capacity of fog node $f$ | Byte | | | |

*\* Memory capacity denotes the amount of RAM, and storage capacity denotes the amount of Hard/SSD drive available to each node.*

$R_{proc}(s)$ and $T^{fog}(s,f,t)$ are measured and provided by the FSP monitoring system. FSP's traffic monitor module monitors the total incoming traffic rate of IoT requests to fog nodes by traffic monitoring agents such as the Software Defined Networking (SDN) controller. Commercial SDN controllers such as OpenDayLight and ONOS have comprehensive traffic-monitoring APIs to monitor application-level traffic. Thus, we have formulated the incoming traffic rate of service $s$ to fog node $f$ as follows:

$$\psi^{fog}(s,f,t) = R_{proc}(s) \, T^{fog}(s,f,t) \qquad (3)$$

---

[3] Million instructions
[4] Million instruction per second

If service $s$ is not deployed on fog node $f$, i.e., $\psi^{fog}(s,f,t) = 0$, it should be deployed on cloud. In this case, the incoming traffic to fog node $f$, i.e., $T^{fog}(s,f,t)$, is rejected by the fog node and is forwarded from the fog node directly to cloud servers, which is equivalent to $\psi^{fog}(s,f,t)$. We have defined the amount of incoming traffic to the cloud server due to offloading the service on a cloud server by the sum of incoming rejected traffic to the cloud server, as shown in (4).

$$\psi^{cloud}(s,k,t) = \sum_{f \in \{f'|h_s(f')=k\}} \left(\psi^{fog}(s,f,t)\right) \quad (4)$$

where the function $h_s(f)$ provides the index of a cloud server that receives the transmission of the traffic of service s from the fog node $f$. The sum operator in (4) is used to aggregate the outgoing traffic from all fog nodes coming to the cloud server, which forms the total incoming traffic to a cloud server.

### 3.1.3 Dedicated processing resources

The amount of processing capacity a server is dedicated to service $s$, i.e., $\Gamma^{fog}(s,f,t)$ and $\Gamma^{cloud}(s,k,t)$, is defined as (5) and (6) for fog and cloud, respectively. These two quantities measure the number of instructions being processed on the server associated with the service $s$.

$$\Gamma^{fog}(s,f,t) = \frac{R_{proc}(s)}{\sum_{s'} P(s',f,t) R_{proc}(s')} M_{proc}^{fog}(f) \quad (5)$$

$$\Gamma^{cloud}(s,k,t) = \frac{R_{proc}(s)}{\sum_{s'} Q(s',k,t) R_{proc}(s')} M_{proc}^{cloud}(k) \quad (6)$$

where $M_{proc}^{fog}(f)$ and $M_{proc}^{cloud}(k)$ denote the processing capacity of the fog and cloud servers, respectively.

### 3.1.4 Service delay

Service delay defines the period between sending the service request and receiving the response by the IoT node. This delay includes propagation delay, the delay associated with waiting in the queue of fog and cloud servers, processing delay, and round-trip transmission delay between IoT and fog nodes. The service delay, $d_{service}(s,f,t)$, is formulated as described in [6].

## 3.2 Constraints

### 3.2.1 Processing, memory, and storage capacities

Each fog node is capable of providing a restricted range of services owing to its finite resources. The number of incoming instructions to a server for service $s$ should not exceed the number of instructions dedicated to this service. This constraint encourages the algorithm to choose a placement that does not violate the capacity limitations formulated in (7) and (8).

$$\psi^{fog}(s,f,t) < \Gamma^{fog}(s,f,t) \quad (7)$$

$$\psi^{cloud}(s,k,t) < \Gamma^{cloud}(s,k,t) \tag{8}$$

In addition to processing resources, memory, and storage resources must be constrained to prevent further resource overload. Therefore, we expect the resource utilization of fog nodes and cloud servers not to exceed the available resources. This condition can be defined as equations 9–12.

$$\sum_s P(s,f,t) R_{stor}(s) < M_{stor}^{fog}(f) \tag{9}$$

$$\sum_s P(s,f,t) R_{mem}(s) < M_{mem}^{fog}(f) \tag{10}$$

$$\sum_s Q(s,k,t) R_{stor}(s) < M_{stor}^{cloud}(k) \tag{11}$$

$$\sum_s Q(s,k,t) R_{mem}(s) < M_{mem}^{cloud}(k) \tag{12}$$

where $M_{stor}^{fog}(f)$ and $M_{stor}^{cloud}(k)$ denote the storage capacity of the fog and cloud servers, respectively. The memory capacity of the fog and cloud servers is also indicated by $M_{mem}^{fog}(f)$ and $M_{mem}^{cloud}(k)$, respectively. Besides, $R_{stor}(s)$ and $R_{mem}(s)$ demonstrate the storage and memory capacities of service, respectively. In the latest equations, the aggregated amount of required memory and storage capacities of services must not exceed the available memory and storage capacities on fog nodes and cloud servers.

### 3.2.2 Release constraints

If there exists at least a single fog node forwarding traffic to cloud servers, *i.e.*, $\sum_f \sum_{k \in \{k'|h_s(f)=k', \forall k' \in C\}} T^{fog}(s,f,t) > 0$, the controller can not release the service from the cloud. This can be formulated as (13).

$$Q(s,k,t) = \begin{cases} 1 & \sum_{f \in \{f'|h_s(f')=k\}} \left(T^{fog}(s,f,t)\right) > 0 \\ 0 & otherwise \end{cases} \tag{13}$$

This means service $s$ can be released from the cloud only if there is no incoming request to the cloud for it. This constraint ensures that release from cloud servers occurs safely. It is worth mentioning that the equations 3–13 are defined as discussed in [6].

### 3.3 Objective statement

In this subsection, the total cost function is formulated w.r.t constraints, and various parts of the cost function are proposed. We have used the term loss for most details of cost from the machine learning convention, where the cost means the average of losses.

### 3.3.1 Cost function

Our objective is focused on maximizing the QoS, in other words, with service delay and delay violation, as well as utilization of servers. We have formulated the total cost function as follows:

$$P1: \text{minimize} \left(\frac{1}{|F| \times |S|}\right)$$

$$\times \left[\sum_{f \in F}\sum_{s \in S}\begin{pmatrix} L_{proc}^{fog} + L_{stor}^{fog} + L_{violation}^{fog} \\ + L_{comm}^{FC}(s,f,h_s(f),t) \\ + L_{dep}^{fog} + L_{wrong}^{fog} + L_{delay}^{fog} \end{pmatrix}\right]$$

$$+ \left(\frac{1}{(|F|)^2 \times |S|}\right) \times \left[\sum_{f \in F}\sum_{f' \in F-\{f\}}\sum_{s \in S} L_{comm}^{FF}(s,f,f',t)\right] \quad (14)$$

$$+ \left(\frac{1}{|F| \times |S|}\right) \times \sum_{f \in F}\sum_{s \in S} L_{utilization}^{fog}(s,f,t)$$

$$+ \left(\frac{1}{|C| \times |S|}\right) \times \left[\sum_{k \in C}\sum_{s \in S}(L_{proc}^{cloud} + L_{stor}^{cloud})\right]$$

where, $|F|$, $|S|$, and $|C|$ show the size of fog nodes, services, and cloud server sets. The first three terms denote losses related to fog nodes, and the rest of the terms denote the losses of cloud servers. In (14), the processing and storage losses in fog and cloud are denoted by $L_{proc}^{fog}$, $L_{stor}^{fog}$, $L_{proc}^{cloud}$, and $L_{stor}^{cloud}$, respectively. $L_{comm}^{FC}$ measures the loss of communication between fog and cloud servers, and $L_{violation}^{fog}$, $L_{dep}^{fog}$, and $L_{delay}^{fog}$ denote delay violation, deployment, and service delay losses in fog, respectively. $L_{utilization}^{fog}$ measures the loss of low resource utilization of fog nodes, and $L_{comm}^{FF}$ denotes the communication loss between fog nodes because of offloading a service on another fog node. All losses are described in [6], except for $L_{utilization}^{fog}$, $L_{dep}^{fog}$, and $L_{delay}^{fog}$.

### 3.3.2 High fog utilization reward

We promote the use of the service provisioning method to discover the most suitable resource for each service in order to attain the most favorable placement, namely, the minimal service delay, along with enhanced resource utilization and QoS. So, we consider the improvement of fog utilization as a reward for the meta-heuristic algorithm [22]. We have considered this reward to encourage the algorithm to utilize fog nodes as much as possible. The utilization of a fog node $f$ is calculated as stated in [23], proposed in (15).

$$L_{utilization}^{fog}(s,f,t) = -\alpha \times \frac{\sum_s P(s,f,t) \times R_{mem}(s)}{M_{mem}^{fog}(f)} - (1-\alpha) \times \frac{\sum_s P(s,f,t) \times R_{stor}(s)}{M_{stor}^{fog}(f)} \quad (15)$$

where $\alpha$ adjusts the balance between resources, known as the "impact coefficient". The impact coefficient is a real number in the range *[0, 1]*. In the above equation, $R_{mem}$ returns the required memory capacity for a specific service. Each fraction in (15) represents the proportion of each resource utilized by the service. As the resource utilization increases, (15) yields a larger value.

*3.3.3    Service delay and deploy delay losses*

We have considered the loss of service delay as the most recent loss. The FSP is penalized for every millisecond of delay, which compels the algorithm to locate the placement that incurs the least service delay. This delay encompasses both service delay and deploy delay, and is formulated as follows:

$$L_{delay}(s,f,t) = \begin{matrix} C_{delay}(f)\, d_{service}(s,f,t)\, \tau + C_{deploy}\left(1 - P(s,f,t-1)\right) \\ \times P(s,f,t)\, d_{deploy}(s,f,t)\, \tau \end{matrix} \qquad (16)$$

where $C_{delay}$ and $C_{deploy}$ represent the costs associated with service and deploy delays per millisecond, respectively. The service delay also has impacts on delay violation. So, the latest loss can lower the delay violation and improve the QoS level of the service as well. In (16), the term $d_{deploy}(s,f,t)$ refers to deploy delay, including download delay of the container from FSC and container startup time. We have defined $d_{deploy}(s,f,t)$ as follows:

$$d_{deploy}(s,f,t) = \frac{R_{stor}(s)}{\xi(f)} + Container\ startup\ delay \qquad (17)$$

Equation (17) uses $\xi(f)$ to represent the transmission rate from FSP's image storage to fog node $f$, measured in bytes per second. The first part of the equation calculates the container download delay from FSP to the server. To maintain the expected QoS, FSC continuously updates placements due to changing traffic conditions. However, P1 is an NP-hard problem, solvable periodically within a reasonable timeframe. Since solving INLP problems in real-time isn't feasible, we've resolved this issue by running the algorithm offline.

**4    Proposed Methodology**

In this section, we delve into the proposed approach, HPCDF, aimed at solving the NP-hard problem of fog service provisioning. HPCDF employs a meta-heuristic approach, exploring the problem's search space with limited iterations. It follows a process shown in Figure 2, where initially, services are randomly placed on fog nodes. Subsequently, a hybrid binary PSO-CRO is executed to optimize service placement, resulting in the best placement. Finally, this optimized placement is mapped back to the fog placement matrix, P.

*4.1    General description of PSO and CRO algorithms*

Particle Swarm Optimization (PSO), developed by Kennedy and Eberhart [17], is a robust meta-heuristic algorithm known for its parameter efficiency and fast convergence. It excels in solving uni-modal problems without local minima, drawing inspiration from the social behaviors of animal swarms like birds, fish, and ants. Chemical Reaction Optimization (CRO), introduced by Lam and Li [9], is an evolutionary meta-heuristic inspired by chemical reactions. CRO is efficient and capable of handling local minima problems effectively through its computational operators.

*4.2  Motivation of hybrid approach*

The PSO algorithm can perform well at exploration, *i.e.*, global search, and the CRO algorithm functions well at exploitation, *i.e.*, local search. On the other hand, PSO would be stuck in local minima when utilized in multi-modal problems. However, an excellent optimization algorithm establishes a balance between exploration and exploitation. The mentioned issues motivated researchers to merge these two algorithms, which resulted in HBPCRO, proposed by Li et al. [18]. Our motivation to utilize HBPCRO is to exploit both algorithms' benefits. HBPCRO must perform periodically to address the optimization problem formulated in the previous section. The way we adapted this algorithm is discussed below.

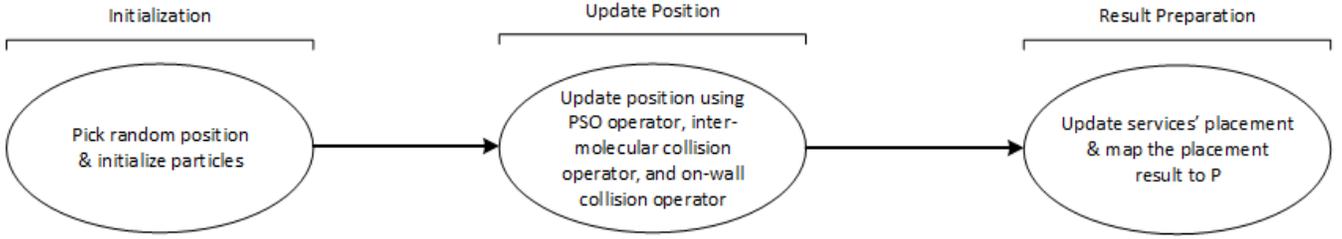

Figure 2. General overview of the proposed approach workflow

*4.2.1  Particle, position, and velocity representation*

Each member of the swarm, a particle[5], investigates in the multi-dimensional search space. The particle's Position denotes a unique placement, *i.e.*, a possible solution. Assume the binary position vector, $X$, is a vector of size $|F| \times |S|$, where

$$X(i,t) = P\left(\left\lfloor \frac{i}{|F|} \right\rfloor, i - |F| \times \left\lfloor \frac{i}{|F|} \right\rfloor, t\right) \tag{18}$$

In (18), $i$, equal to $s \times |F| + f$ denotes the dimension index, where $s$ and $f$ represent the service index and fog node index, respectively. $X(i,t) \in \{0,1\}$, which signifies the i-th element of particle position, denotes whether a service is deployed on a fog node or not. The service index is determined by $\left\lfloor \frac{i}{|F|} \right\rfloor$, while the fog node index is calculated by $i - |F| \times \left\lfloor \frac{i}{|F|} \right\rfloor$.

The position vector and its equivalent matrix are shown in Figure 3. The placement matrix (P), drawn in Figure 3 (a), defines each service's placement on each fog node. Figure 3 (b) shows the flat form of the placement matrix where the whole rows are arranged one after another. The HPCDF approach chooses random values as the initial position, *i.e.*, placement, for all particles.

The velocity of a particle, which determines the speed of a particle in search space, is a vector of $|F| \times |S|$ real numbers, *i.e.*, $V(i,t) \in \mathbb{R}$. In other words, velocity $V(i,t)$ determines how probable the placement of service indexed as $\left\lfloor \frac{i}{|F|} \right\rfloor$ on the fog node with the index of $i - |F| \times \left\lfloor \frac{i}{|F|} \right\rfloor$. It is worth mentioning that, in Figure 3 (a), the fog placement matrix is a sparse binary matrix,

---

[5]We have assumed a "particle" in PSO is equal to a "molecule" in CRO, and the word "particle" is used in the rest of this paper.

where each row denotes a service, and each column denotes a fog node. A value of 1 means the service should be deployed on the corresponding fog node. On the other hand, in Figure 3 (b), a flattened vector of the placement matrix is denoted. Values of this flattened vector are updated by the proposed method.

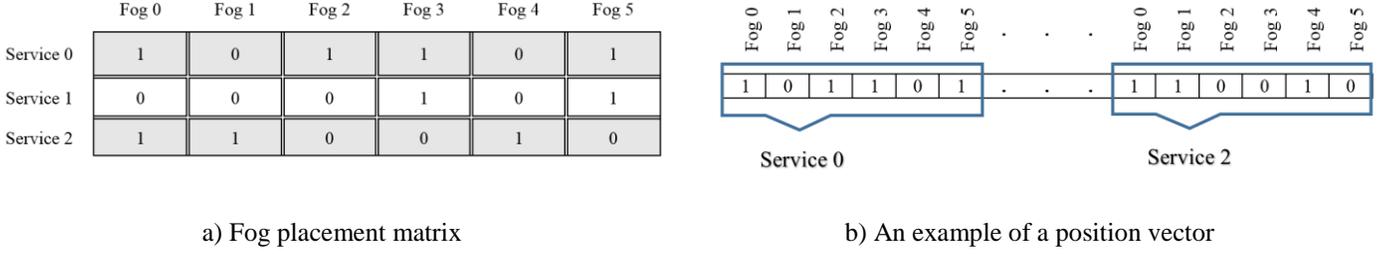

a) Fog placement matrix          b) An example of a position vector

Figure 3. Services placement matrix and position vector

### 4.2.2 Particle's movement

Each particle updates its position according to two values, local best ($p_{best}$) and global best ($g_{best}$), in each iteration of PSO. $p_{best}$ is the position with the least cost inside each particle. On the other hand, $g_{best}$ is the most minor cost position among all particles inside the swarm. We have to update $p_{best}$, $g_{best}$, velocity, and position periodically. The way we update velocity and position is discussed further. In the original version of PSO [17], $V(i, t+1)$ is calculated using (19).

$$V(i, t+1) = \omega V(i,t) + \alpha U(0,1)\left(p_{best} - X(i,t)\right) + \beta U(0,1)\left(g_{best} - X(i,t)\right) \tag{19}$$

In the latest equation, $\omega$, $\alpha$, and $\beta$ are inertia, cognitive, and social coefficients, respectively. $\alpha$ and $\beta$ are called acceleration coefficients, conversely. $U(0,1)$ is a sample of a uniform random number in the range *[0, 1]*, diversifying the particle's motion in various directions. In (19), $\omega$, $\alpha$, and $\beta$ are problem-dependent and should be adjusted for the particular problem. It is proved that changing the hyper-parameters of PSO results in better solutions.

On the other hand, to accelerate the algorithm, one can achieve a fast convergence version of PSO by changing the hyper-parameters according to the distribution of particles. For this reason, a *spread* factor is utilized, as stated in [8], to modify the value of inertia weight continuously. The *spread* is defined as follows,

$$spread = \frac{(precision + deviation)}{2} \tag{20}$$

where *precision* is defined as the greatest disparity in fitness value between particles, while *deviation* is the Euclidean distance between the position of the best particle in the global and the average position of all particles. The spread factor is utilized to determine the inertia weight through the use of the following equation.

$$\omega = \exp\left(\frac{-iteration\_num}{spread \times MaxIter}\right) \tag{21}$$

where *iteration_num* denotes the current iteration, and *MaxIter* shows the maximum iteration of the algorithm.

On account of changes in the number of fog nodes and services over time in the fog service provisioning problem, the mentioned hyper-parameters must be adjusted dynamically. To update the velocity, we have implemented the approach that was discussed in [19]. The way for updating these hyper-parameters is formulated as (22) and (23).

$$\alpha = (\alpha_f - \alpha_i) \times \left(\frac{iteration\_num}{MaxIter}\right) + \alpha_i \tag{22}$$

$$\beta = (\beta_f - \beta_i) \times \left(\frac{iteration\_num}{MaxIter}\right) + \beta_i \tag{23}$$

where $\alpha_i$ and $\beta_i$ are initial values, and $\alpha_f$ and $\beta_f$ are final values of hyper-parameters $\alpha$ and $\beta$, respectively. $\alpha$ and $\beta$ are updated according to iterations. To avoid the occurrence of particle fluctuations and abrupt movements of particles between different regions in the search space, we have implemented a method known as velocity clipping. This technique controls the values of the velocity vector according to (24).

$$V(i,t) = \begin{cases} V_{min} & V(i,t) < V_{min} \\ V(i,t) & V_{min} < V(i,t) < V_{max} \\ V_{max} & V(i,t) > V_{max} \end{cases} \tag{24}$$

where $V_{min}$ and $V_{max}$ are the minimum and maximum values of velocity, respectively.

Apart from this, the original version of PSO is defined for continuous problems. During the provisioning of services, as well as in the overall scheduling of tasks, there exists a discrete problem. For this purpose, we have substituted the original version of PSO with a binary version proposed by Kennedy and Eberhart [17], where they differ in how the position is updated. Hence, we have used the sigmoid function to translate velocity to probability in the range *[0, 1]*, which determines to deploy, *i.e.*, 1, or to release, *i.e.*, 0, the service on fog nodes. The sigmoid function is defined as follows,

$$\sigma(z) = \frac{1}{(1 + e^{-z})} \tag{25}$$

The way we interpret a particle's velocity is discussed in [30] and is as follows. The more positive the velocity becomes, the more likelihood of deploying the service to decide whether to deploy or to release the service. On the other hand, the more negative the velocity becomes, the more likely it is to release the service. This condition compels the transfer function to transform into a V-shaped function [20]. Equation (26) shows how to convert the sigmoid function to a V-shaped function and update the position vector elements.

$$X(i,t) = \begin{cases} 1 & (V(i,t) > 0) \text{ and } (|2 \times (\sigma(V(i,t)) - 0.5)| \geq randu) \\ 0 & (V(i,t) \leq 0) \text{ and } (|2 \times (\sigma(V(i,t)) - 0.5)| \geq randu) \\ X(i,t-1) & |2 \times (\sigma(V(i,t)) - 0.5)| < randu \end{cases} \tag{26}$$

where $randu$ denotes a uniform random number that is sampled per dimension. Figure 4 shows the sigmoid function's behavior according to input values and denotes how we have modified the sigmoid function using (26), which ignores the sign and only considers the magnitude of the input value.

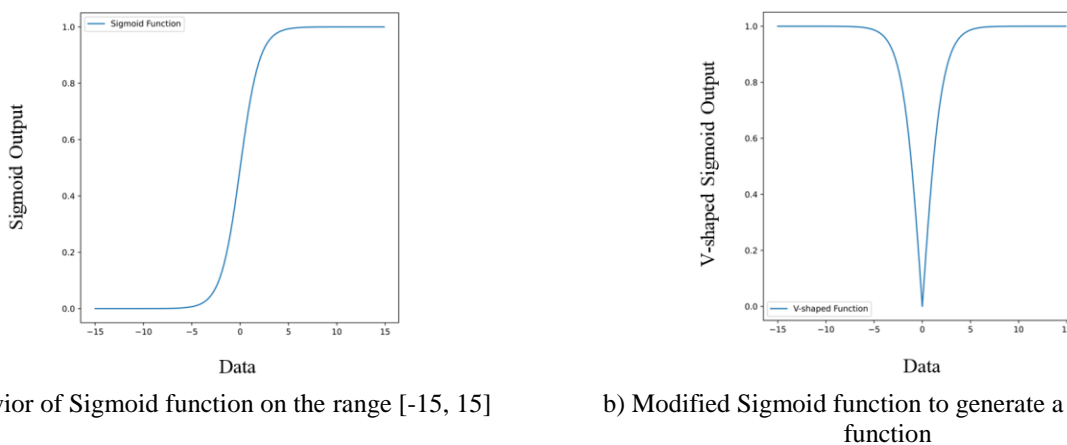

a) Behavior of Sigmoid function on the range [-15, 15]       b) Modified Sigmoid function to generate a V-shaped function

Figure 4. Comparison between Sigmoid and Modified Sigmoid functions

It is important to note that, in Figure 4 (a), the sigmoid function is represented as the original one. This function is an S-shaped function, where its domain lies between 0 and 1 and can be interpreted as probability. In Figure 4 (b), the sigmoid is reformed to denote a V-shaped function, where a very positive or very negative input value results in the deployment or release of the service, respectively.

Algorithm 1 outlines the execution of the $updatePosition$ operator in HBPCRO for particle position updates. It starts with an attempt to place services on fog nodes, checking if the velocity aligns with the right side of a V-shaped function, indicating possible fog node deployment (Lines 7–14). Similarly, the left side of the V-shaped function signifies the likelihood of service release from fog nodes. Deployed services are then released from cloud servers and vice versa (Lines 15–21). After each decision, the service delay violation for the new placement is computed.

The CRO operators are executed probabilistically, considering potential and kinetic energies. Potential energy reflects the fitness function or cost in the current position, while kinetic energy enables particles to move to positions with higher potential energy, aiding in escaping local minima. The $onWallCollision$ operator, detailed in Algorithm 2, handles instances where particles encounter walls, resulting in slight adjustments to particle structure, including position and potential energy. The algorithm involves selecting two random indices from a position vector and swapping the values at those indices (Lines 4–5), ensuring constraints are met (Line 6), and invoking a best-fit algorithm if needed [21]. If placement remains unsuccessful, the process is repeated until constraints are satisfied.

| **Algorithm 1** updatePosition | **Algorithm 2** onWallCollisionOperator |

<table>
<tr><td>

1: **Input**: *S, F, C, Velocity vector V*
2: **Output**: *Updated placement matrices P and Q*
3: **for** *each service s* **do**
4:   **for** *each fog node f* **do**
5:     $randNum \leftarrow U(0,1)$
6:     $V_{s_f} \leftarrow V(s \times |F| + f, t)$
7:     **if** $V_{s_f} > 0$ and $|2 \times (\sigma(V_{s_f}) - 0.5)| >= randNum$ **then**
8:       **if** *f has free resources* **then**
9:         deploy service s on fog node f
10:         recalculate violation
11:     **else**
12:       **if** $V_{s_f} \leq 0$ and $|2 \times (\sigma(V_{s_f}) - 0.5)| >= randNum$ **then**
13:         release service s from fog node f
14:         recalculate violation
15:   **for** *each cloud server k* **do**
16:     **if** the *traffic of service s is forwarded to cloud server k* **then**
17:       deploy service s on cloud server k
18:       recalculate violation
19:     **else**
20:       release service s from cloud server k
21:       recalculate violation

</td><td>

1: **Input**: *S, F, selected particle, localThresh, constraints, MinKeLossPer, energy*
2: Copy the selected particle into a new particle
3: **for** *the limited number of steps* **do**
4:   pick random numbers i and j in the range [0, |F|×|S| - 1]
5:   swap i-th and j-th elements of new particle
6:   **if** *constraints are violated* **then** bestFit(.) is invoked
7: Calculate the potential energy of the new particle based on the cost value
8: **if** *potential energy is decreased through collision* **then**
9:   pick a random number q in the range [MinKeLossPer, 1]
10:   keep q% of lost energy as new kinetic energy
11:   update personal best position and cost
12:   old particle ← new particle
13:   localThresh += 1
14:   update the global best position

</td></tr>
</table>

Potential energy is recalculated (Line 7), and if it decreases, kinetic energy is updated, and the old particle is replaced (Lines 8–12). The new particle's *localThresh* attribute is incremented (Line 13) to signify an additional local search step. The *interactMolecularCollision* operator, described in Algorithm 3, represents instances where two particles collide, resulting in new molecules with altered positions and potential energies.

Algorithm 3 builds on Algorithm 2, performing a swap between two different particles. Two random elements from the position vectors of these particles are swapped crosswise (Lines 3–12). Constraint checks are then performed, invoking the best-fit algorithm if violations occur. Subsequently, the combined potential energy of both particles is calculated (Line 13), and if it's reduced, kinetic energies are updated based on energy loss. The old particles are replaced with the new ones (Lines 14–20), and the *localThresh* attribute of the new particles is incremented (Lines 21–22). These operators prevent the algorithm from getting stuck in local minimum positions due to abrupt position changes.

### 4.2.3 Initial population

The initial position of particles is populated with binary random values, and velocity values are sampled from the random uniform distribution. It is worth mentioning that the initial position is considered as initial $p_{best}$, and the cost associated with it is equal to positive infinity.

*4.2.4    The manipulated hybrid PSO-CRO algorithm*

We have proposed the hybrid binary PSO-CRO algorithm in Algorithm 4 to provision services in the hybrid fog-cloud environment to minimize the service delay and QoS violation.

| **Algorithm 3** interMolecularCollisionOperator | **Algorithm 4** hybridBinaryPsoCro (HBPCRO) |
|---|---|
| 1: **Input**: S, F, selected particles, localThresh of both selected particles, constraints, energies of both selected particles | 1: **Input**: *Set of particles, iterations, interProb, γ* |
| 2: Copy the chosen particles into two new particles | 2: **Output**: *Updated placement matrix P* |
| 3: **for** *the limited number of steps* **do** | 3: initialize particles set |
| 4:     pick random numbers i and j in range [0, \|F\|×\|S\| - 1] | 4: **for** *each particle* **do** |
| 5:     swap i-th elements of the first new particle and the first old particle | 5:     update personal best position, cost, and $X_{opt}$ |
| 6:     swap j-th elements of the first new particle and the second old particle | 6: **for** *a limited number of iterations* **do** |
| 7:     **if** *constraints are violated,* **then** revert changes | 7:     pick a random particle p |
| 8: **for** *the limited number of steps* **do** | 8:     **if** *p.localThresh > γ* **then** |
| 9:     pick random numbers i and j in range [0, \|F\|×\|S\| - 1] | 9:         **for** *each particle* **do** |
| 10:    swap i-th elements of the second new particle and the first old particle | 10:            update velocity V and positions X through the updatePosition(.) operator |
| 11:    swap j-th elements of the second new particle and the second old particle | 11:            update personal best position, cost, and $X_{opt}$ |
| 12:    **if** *constraints are violated,* **then** revert changes | 12:        p.localThresh = 0 |
| 13: Calculate the potential energy of new particles based on cost value | 13:    **else** |
| 14: **if** *potential energy is decreased through collision,* **then** | 14:        **if** *interProb < U(0, 1)* **then** |
| 15:    pick a random uniform number q | 15:            select another random particle p′ |
| 16:    keep q% of lost energy as the first new particle's kinetic energy | 16:            perform interMolecularCollisionOperator(.) |
| 17:    keep (1-q)% of lost energy as the second new particle's kinetic energy | 17:        **else** |
| 18:    update personal best positions and costs | 18:            perform onWallCollisionOperator(.) |
| 19:    first old particle ← first new particle | 19:        confirm position consistency |
| 20:    second old particle ← second new particle | 20:    update $X_{opt}$ among the swarm |
| 21:    first new particle's localThresh += 1 | 21:    **if** *violation and delay thresholds are satisfied* **then** |
| 22:    second new particle's localThresh += 1 | 22:        break the iterations loop |
| 23:    update the global best position | 23: Translate the position vector of $X_{opt}$ back to placement matrix P |

We first initialize the swarm of particles (Lines 3–5), and the initial $p_{best}$ of all particles and global $g_{best}$ are determined. Then, the attribute *localThresh*, which denotes the number of times that CRO operators are executed inside a random particle, is compared with the hyper-parameter $\gamma$ (Line 8). $\gamma$ determines the maximum times that particle can investigate in the neighborhood, and the comparison reveals it is time to execute a global search through performing PSO. Afterwards, particles' $p_{best}$ and global $g_{best}$ are updated in the style of PSO, and then, reset the attribute *localThresh* to perform the local search again in $\gamma$ future iterations (Lines 9–12). The global best particle, denoted as $X_{opt}$ in Algorithm 4, represents the particle with the optimal position and cost globally. Lastly, the hyper-parameter *interProb* is compared to a randomly generated uniform number (Line 14), which selects each CRO operator in each iteration randomly.

Whether the $onWallCollision$ operator or the $interMolecularCollision$ operator is executed, the position values in various dimensions are changed randomly. So, we must confirm that the resulting position values are consistent and the constraints are satisfied (Line 19).

*4.3    Time complexity*

The time complexity of the hybrid binary PSO-CRO enormously depends on how PSO and CRO operators are implemented. The initialization step's complexity is $O(|F| \times |S| \times N)$, where $N$ denotes the number of particles. The initializer function has to iterate through the position and velocity vectors to initialize them for $N$ times. The time complexity of executing PSO is $O(|F| \times |S| \times N)$. If CRO operators are executed, the complexity becomes $O(|F| \times |S|)$ because it has to iterate through the position vector and update its values. Finally, the time complexity of the proposed approach is $O(|F| \times |S| \times N)$, which can be approximated to $O(N^3)$.

## 5    Experimental Result

In this section, we detail our experimental setup, analyze results, and evaluate our approach's optimality. We used the FogPlan framework in Java to simulate the Fog-Cloud Datacenter, running experiments on an Intel® Core™ i5-2430M CPU @ 2.40GHz ×4. Three experiments explored hyper-parameter influence on placement optimality in HPCDF: (1) optimizing delay and QoS through tuning the parameters of hybrid binary PSO-CRO, (2) comparing our approach to greedy baselines in terms of service delay, violation, and cost, and (3) studying delay variation's impact on placement optimality. Subsequent subsections provide experiment details and results analysis.

*5.1    Experiment setup*

*5.1.1    Network topology*

Fog nodes are initially randomly connected to cloud servers, and these connections are changed over time w.r.t services' placement. As stated before, we take the aggregated incoming traffic of all services from all IoT nodes. Accordingly, there is no need to have information about the number of IoT nodes. Table 3 shows the number of machines and services used in all experiments.

Table 3. Number of fog nodes, cloud servers, and services for all experiments

| Experiment | Num. of fog nodes | Num. of cloud servers | Num. of services |
|---|---|---|---|
| 1 | 10 | 3 | 40 |
| 2 | 10 | 5 | 50 |
| 3 | 10 | 5 | 20 |

*5.1.2    Mobile Augmented Reality (MAR) application services*

We have considered MAR services in our experiments. MAR refers to a category of ARs which you can have everywhere. The size of request and response messages in this application is in the range of 10 KB to 26 KB and 10 Bytes to 20 Bytes, respectively [22]. We have sampled the required processing capacity of these services from $U(50, 200)$[6] MI per request [23].

*5.1.3    Parameters and delays*

We have sampled the QoS level of each service from $U(0.8, 0.99)$, where 0.99 denotes a strict QoS level. The delay threshold of each service is sampled from $U(10, 15)$ ms [24]. The propagation delay between IoT nodes and fog nodes is sampled from $U(1, 2)$ ms, and between fog nodes and cloud servers is sampled from $U(15, 35)$ ms [25]. We have assumed between 6 to 10 hops distance between fog nodes to cloud servers, with a maximum of two 100 Gbps links, and other links would be 10 Gbps. The transmission rate between IoT nodes to FSC is assumed to be 10 Gbps. The deployment and release delays are less than 50 ms [26] and may be ignored. However, we have considered it in the objective in Section 3. Delay of deployment includes the time needed to download containers from FSC onto the machine, and the time required for container startup is considered equal to 50 ms. It is worth mentioning that we have set the impact coefficient, $\alpha$, equivalent to 0.5. The hyper-parameters of the hybrid binary PSO-CRO are shown in Table 4. These parameter values are determined using experiment 1, which is investigated in the following subsection.

Table 4. Experiments' configuration for HPCDF

| Hyper-parameter | Value |
| --- | --- |
| Number of particles | 35 |
| Iterations | 700 |
| Gamma ($\gamma$) | 3 |
| Initial KE | 100000 |
| Velocity range | [-0.5, 5.0] |
| minKElossPer | 0.1 |
| interProb | 0.8 |
| $\alpha$ | Linearly decrease from 0.9 to 0.5 |
| $\beta$ | Linearly increase from 0.5 to 5.5 |

*5.1.4    Services and machines' capacities*

In our experiments, we have considered heterogeneous machines to show their impact on the placement. The processing capacity of each fog node is sampled from $U(800, 1300)$ MIPS, and the processing capacity of each cloud server is sampled from $U(16000, 26000)$ MIPS. Storage capacity for each fog node and the cloud server is between 10 GB to 25 GB and 100 GB to 250

---
[6] $U(i, j)$ denotes a normal random number from the range of [i, j]

GB, respectively. Memory capacity for each fog node and the cloud server is between 4 GB to 16 GB and 8 GB to 32 GB, respectively. Some instance types of Amazon EC2 machines[7] are listed in Table 5, suitable for fog nodes and cloud servers.

Table 5. Amazon EC2 instance types

| **Fog node** | | | **Cloud server** | | |
|---|---|---|---|---|---|
| Instance type | Num. of cores | RAM (GB) | Instance type | Num. of cores | RAM (GB) |
| A1 XL | 4 | 8 | A1 XXL | 8 | 16 |
| T3 XL | 4 | 16 | T3 XXL | 8 | 32 |
| T2 XL | 4 | 16 | T2 XXL | 8 | 32 |
| M5 XL | 4 | 16 | M5 XXL | 8 | 32 |
| M5d XL | 4 | 16 | M5d XXL | 8 | 32 |
| M5a XL | 4 | 16 | M5a XXL | 8 | 32 |
| M5ad XL | 4 | 16 | M5ad XXL | 8 | 32 |
| M4 XL | 4 | 16 | M4 XXL | 8 | 32 |

*5.1.5 Costs*

The processing cost in a fog node or a cloud server equals $2 \times 10^{-3}$ per million instructions. The storage cost in a fog node or a cloud server equals $4 \times 10^{-3}$ per B/s and 4 per B/s, respectively. The communications between fog nodes and cloud servers and between fog nodes and FSC cost 0.2 per Gb and 0.5 per Gb, respectively. The cost of a 1% violation is considered between 100 to 200 per second. Also, the cost associated with the release of requests from nodes experiencing high traffic is considered to be 2 per request per second. Additionally, the cost related to service delay is equivalent to $4 \times 10^{-3}$ per millisecond.

*5.1.6 Dataset*

To experiment with the proposed approach in a reasonably realistic environment, we have used real-world traffic traces provided by the MAWI working group [27]. This dataset is collected from project WIDE, where research is performed on computer networks. MAWI datasets consist of PCAP files, utilizing the tcpdump format, albeit lacking the inclusion of the payload. Traffic traces are made by tcpdump, and then IP addresses in the traces are scrambled by a modified version of tcpdpriv. Traces of five sample points are available. The traffic data that is intended for analysis is chosen from SamplePoint-F, specifically on the 12th of April 2017, between the hours of 12:00 PM and 2:00 PM. These data represent the incoming traffic directed towards the fog nodes. We have extracted information from traced TCP flows and aggregated these statistics by geographical location and network prefix. The traffic in this dataset changes every minute. We have used ten subnets of the largest traffic machines as fog nodes.

---

[7] https://aws.amazon.com/ec2/instance-types

## 5.2 Numerical Results

In this subsection, we shall examine the impact of the delay threshold on the optimality of the proposed algorithm. The proposed algorithm is compared with baselines, like All-Cloud, Min-Viol, Min-Cost, and BPSO. In All-Cloud, all services are provisioned on cloud servers. Min-Cost and Min-Viol, as greedy methods, provision services to decrease delay violation and service cost, respectively. BPSO improves the services' placement by using pure binary PSO. All experiments are discussed below.

### 5.2.1 Experiment 1

In the first experiment, we explored the impact of the number of particles and the hyper-parameter gamma, *i.e.*, $\gamma$, on service delay, delay violation, and service cost in our proposed approach. This experiment involved testing the ranges *[2, 10]* and *[5, 50]* for gamma values and number of particles through random searches over 637 iterations. The results are presented in Figure 5, where the x-axis represents delay violation, the y-axis represents the number of particles, and the z-axis represents gamma values. Service delay and cost are indicated by point size and color. To enhance clarity, service delay and cost values were scaled to the range [20, 80]. The findings in Figure 5 indicate that increasing the number of particles and reducing $\gamma$ lead to decreased service cost, service delay, and delay violation. Consequently, Table 4 sets the number of particles and gamma values based on these results.

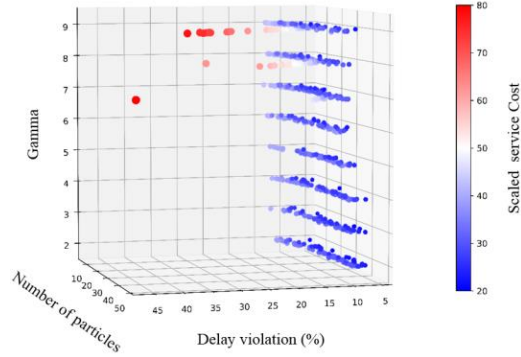

Figure 5. Impact of hyper-parameters gamma and number of particles on service delay, delay violation, and service cost

### 5.2.2 Experiment 2

In the second experiment, we investigate the optimality of HPCDF compared to All-Cloud, Min-Cost, and Min-Viol. The result of this experiment is an average of 10 consecutive runs of the experiment. In this experiment, traffic is changed every 60 seconds, and we have set $\tau$ equal to 120 seconds. Figure 6 (a) reveals the normalized traffic from 12:00 p.m. to 2:00 p.m.

Figure 6 (b) displays average service delay results, showcasing the proposed method's 66.02% improvement in service cost compared to Min-Viol. The highest service cost is observed in the All-Cloud approach due to strict QoS on cloud servers, leading to delay violations and cost escalation. In contrast, HPCDF demonstrates greater cost stability. Figure 6(c) illustrates the average service delay, with All-Cloud exhibiting the highest delay due to the deployment of services with strict QoS-level on cloud servers.

HPCDF enhances average service delay by 29.34% compared to Min-Viol, whereas BPSO struggles with local minima issues, which HPCDF overcomes through a balanced exploration-exploitation approach using CRO.

Figure 6 (d) represents the average delay violation, with All-Cloud having the highest violation. Min-Cost exhibits instability with variations in service delay and delay violation. HPCDF increases average delay violation by 50.15% as it focuses on enhancing service cost while not directly considering violation. Figures 6 (e) and 6 (f) show service deployments on cloud and fog servers. HPCDF deploys more services on fog nodes than on cloud servers, resulting in improvements in service delay, cost, and violation. Due to limited fog node resources, deploying the entire services there is impractical, but this approach strikes a balance.

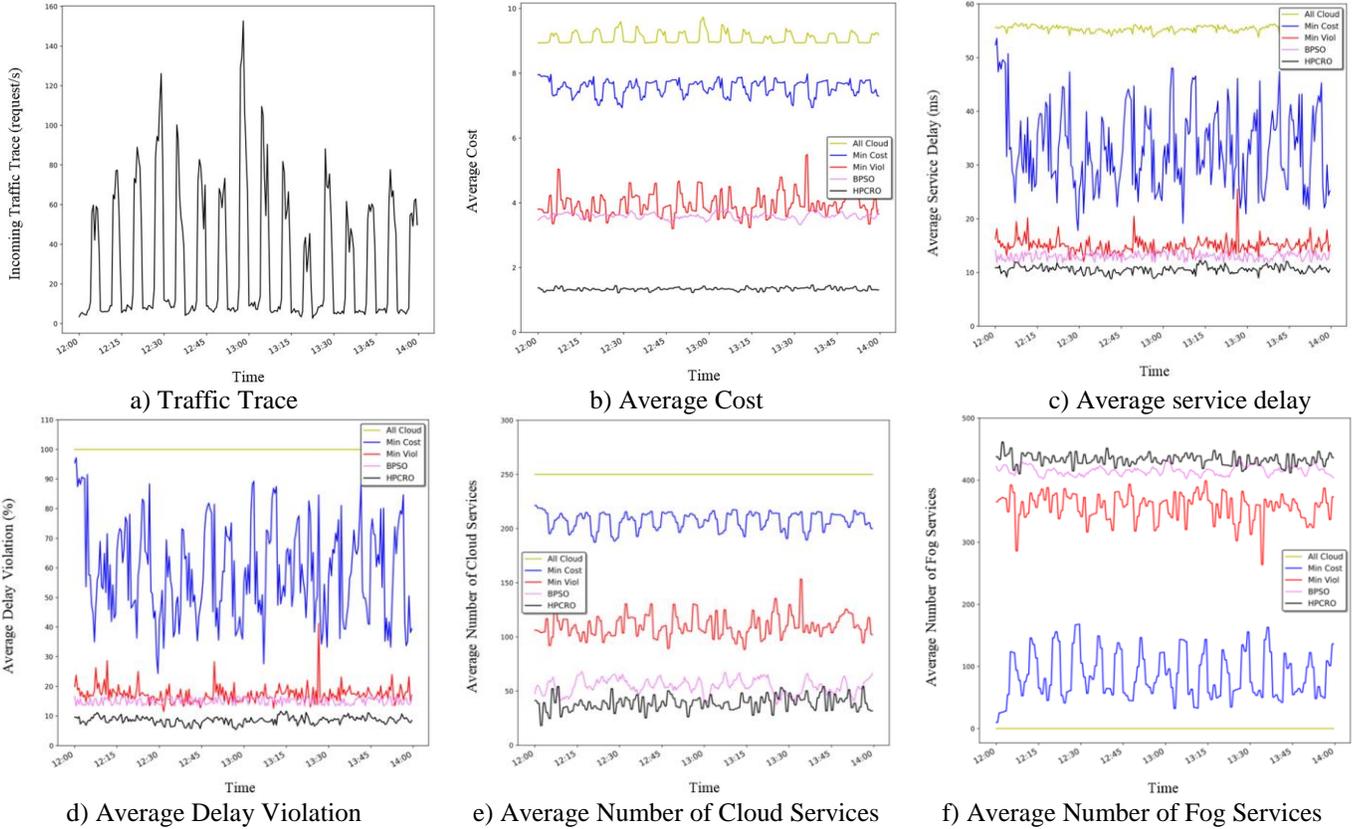

a) Traffic Trace  b) Average Cost  c) Average service delay

d) Average Delay Violation  e) Average Number of Cloud Services  f) Average Number of Fog Services

Figure 6. Simulation result of the second experiment

### 5.2.3 Experiment 3

In the third experiment, we examined the influence of delay thresholds on service delay, cost, violation, and the deployment of services on fog and cloud servers. The results represent the average of 10 consecutive runs and involve changing traffic every 10 seconds with $\tau$ set to 10 seconds. The delay threshold ranged from 1 to 100 milliseconds, and the findings are presented in Figure 7. Figure 7 (a) demonstrates that when using HPCDF, service cost decreases as the delay threshold increases until reaching 11 milliseconds. Beyond 11 milliseconds, service cost gradually rises, with an explosive increase between 66 and 76 milliseconds before stabilizing. Higher delay thresholds lead to increased service deployment on cloud servers, raising service costs. For thresholds above 81 milliseconds, a constant number of services are deployed on both fog and cloud servers due to fog node utilization and cost considerations. This results in a steady service cost for HPCDF compared to Min-Viol, where the delayed increase in fog node service deployment leads to higher costs. Increasing the delay threshold generally leads to deploying more services on cloud servers and higher service costs.

In Figure 7(b), service delay gradually increases with delay thresholds below 66 milliseconds. However, between 66 and 81 milliseconds, service delay sharply rises due to increased deployment on cloud servers. Figure 7(c) reveals that delay violation reduces for thresholds below 11 milliseconds and disappears beyond that, with decreasing QoS sensitivity and delay violation as thresholds rise. Figures 7(d) and 7(e) reveal that higher thresholds result in more services on cloud servers due to reduced QoS sensitivity. As cloud deployments increase, fewer services are placed on fog nodes, ensuring resource stability. Beyond 81 milliseconds, a constant number of services are deployed on fog nodes for improved resource utilization.

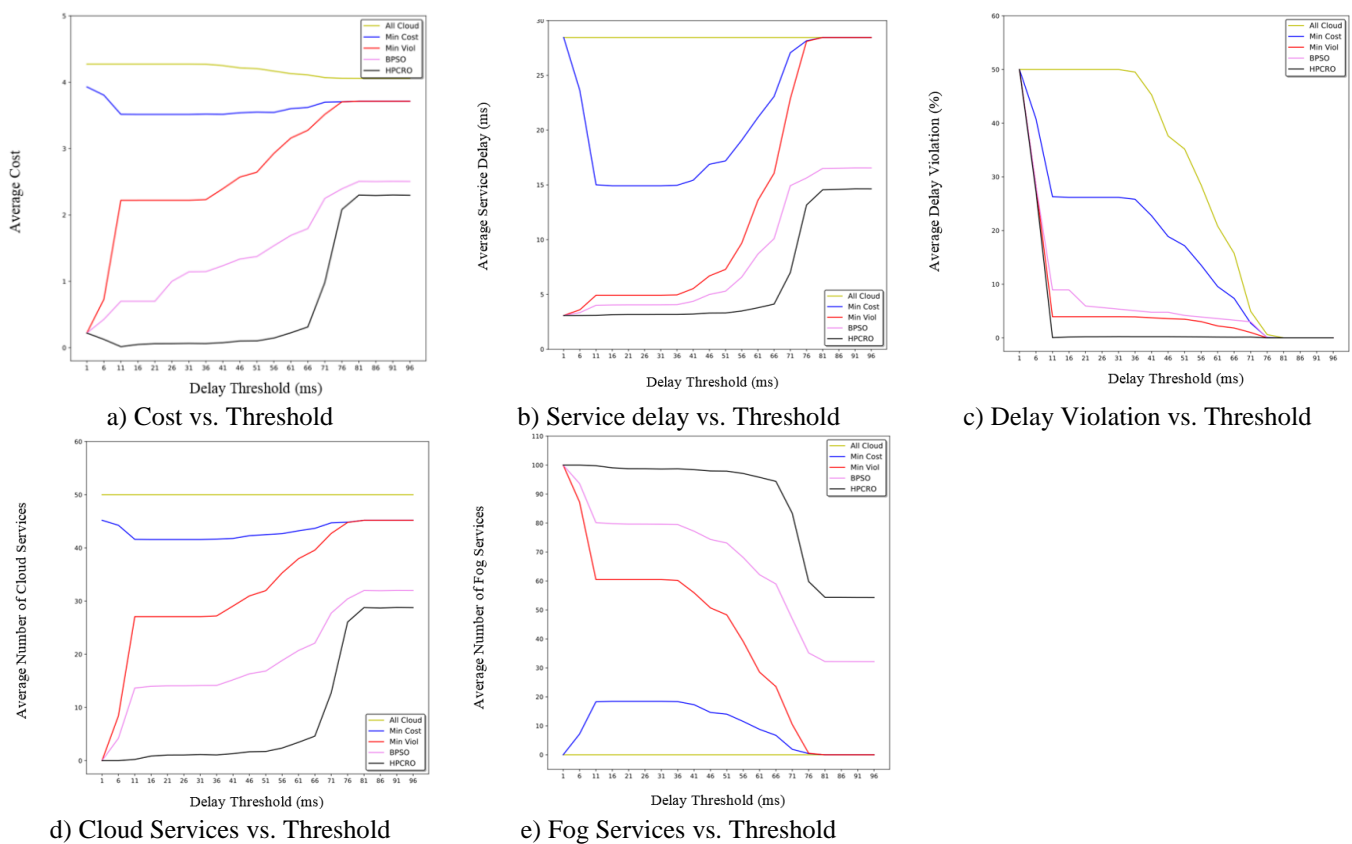

a) Cost vs. Threshold
b) Service delay vs. Threshold
c) Delay Violation vs. Threshold
d) Cloud Services vs. Threshold
e) Fog Services vs. Threshold

Figure 7. Effect of delay threshold on other parameters

## 6  Discussion

The results of experiments indicate that hybrid binary PSO-CRO can achieve a more optimal solution than greedy approaches, but the proposed approach is slower than greedy ones. As can be seen in experiments, the proposed meta-heuristic approach achieves lower service, service delay, and delay violation than greedy strategies. It is evident from experiments that reducing the reconfiguration interval decreases intended parameters slightly. Using a learning algorithm, it is possible to estimate the reconfiguration interval in order to further decrease service delay and delay violation.

HPCDF faces limitations related to slowness and the use of a fixed set of hyper-parameters for incoming traffic. To address these issues, improvements can be made by enhancing the convergence speed of hybrid binary PSO-CRO for high-dimensional search spaces, as discussed in [28-30]. Additionally, a learning strategy can be employed to estimate unique hyper-parameter sets for each incoming traffic scenario, making the algorithm more customized and adaptable to the current traffic conditions.

# 7  Conclusion

Fog computing addresses issues like latency, bandwidth, energy, and QoS for delay-sensitive applications. We introduced an optimization problem and HBPCRO, a hybrid meta-heuristic, to tackle it. Real-world experiments showcased HBPCRO's superior performance, minimizing service delay, violations, and costs compared to greedy methods. Reducing reconfiguration intervals slightly improved service delay and violations, while increasing the delay threshold affected all parameters.

For future works, the proposed algorithm can be adapted for stateful service provisioning, including migrating CPU, memory, hard disk, and network configurations. It can also accommodate flexible resources like Docker containers, prioritizing resource adjustments over deploying new instances, reducing cost and migration delay. Additionally, optimizing placement based on server availability and considering real-time scheduling of services can make the approach more realistic and industry-applicable. Predicting and storing placements for anticipated requests in a database can enable real-time deployment, enhancing the method's usability in production scenarios.

**Soroush Hashemifar** received his B.Sc. degree in computer engineering from Razi University of Kermanshah, in 2020. He was part of High Performance Computing (HPC) lab as a research assistant until the beginning of 2021. His research interests include deep neural networks, computer vision, software engineering, and cloud computing.

**Amir Rajabzadeh** received his Ph.D. degrees in computer engineering from Sharif University of Technology, Iran, in 2005. He has been working as an assistant professor of Computer Engineering at Razi University, Kermanshah, Iran since 2005. He is Dean of innovation science and technology faculty since 2018. His main areas of interests are computer architecture, IoT, and fault-tolerant systems design.